\documentclass[twocolumn,aps,prl,superscriptaddress,nofootinbib,showpacs]{revtex4-1}
\usepackage{graphicx}
\usepackage{color}
\graphicspath{{figures/}{fig/}}
\usepackage{amsmath}
\usepackage{bm}
\usepackage{slashed}
\usepackage{epsfig}
\usepackage{amsfonts}
\usepackage{epstopdf}
\usepackage{hyperref}
\usepackage{bbm}
\usepackage{textcomp}
\usepackage{multirow}

\newcommand{\sect}[1]{{\it \textbf{#1} --- }}
\newcommand{\ui}{\mathrm{i}}
\newcommand{\ud}{\mathrm{d}}
\begin{document}
\title{Multiloop corrections for collider processes using auxiliary mass flow}

\author{Xiao Liu}
\email{xiao6@pku.edu.cn}
\affiliation{School of Physics and State Key Laboratory of Nuclear Physics and
Technology, Peking University, Beijing 100871, China}
\author{Yan-Qing Ma}
\email{yqma@pku.edu.cn}
\affiliation{School of Physics and State Key Laboratory of Nuclear Physics and
Technology, Peking University, Beijing 100871, China}
\affiliation{Center for High Energy Physics, Peking University, Beijing 100871, China}
\affiliation{Collaborative Innovation Center of Quantum Matter,
Beijing 100871, China}

\date{\today}

\begin{abstract}
With a key improvement, the auxiliary mass flow method is now able to compute Feynman integrals encountered in cutting-edge collider processes.
We have successfully applied it to compute some integrals involved in two-loop electroweak corrections to $e^+e^-\to HZ$, two-loop QCD corrections to $3j$, $W/Z/H+2j$, $t\bar{t}H$ and $4j$ production at hadron colliders, and three-loop QCD corrections to $t\bar{t}$ production at hadron colliders, all of which are crucial for precision frontier in collider physics in the following decade. Our results are important building blocks and benchmarks for future studies of these processes.
\end{abstract}

\maketitle
\allowdisplaybreaks

\sect{Introduction}
With the good performance of the Large Hadron Collider (LHC), the particle physics now enters the era of precision. By comparing high-precision experimental measurements with theoretical predictions, we can deepen our understanding of particle physics Standard Model and probe signals of new physics. With the accumulation of experimental data at the LHC and proposals of future colliders, the precision of many experiments will surpass the current theoretical predictions~\cite{Cepeda:2019klc,CEPCStudyGroup:2018ghi}.
Therefore, one of the most important tasks for theorists is to compute higher-order corrections in perturbation theory to reduce theoretical uncertainties for important processes, such as two-loop electroweak corrections to $HZ$ production at $e^+ e^-$ colliders, two-loop QCD corrections to $3j$, $W/Z/H+2j$, $t\bar{t}H$, $4j$ production and three-loop QCD corrections to $t\bar{t}$ production at hadron colliders, and so on, where $j$ means jet.

To carry out these urgently needed perturbative calculations, an important part is to calculate corresponding Feynman master integrals (MIs), which form a complete basis of general Feynman integrals. Although there have been many methods on the market, such as traditional differential equations method which sets up and solves differential equations (DEs) with respect to kinematical variables $\vec{s}$ (denoted as $\vec{s}$-DEs)~\cite{Kotikov:1990kg, Kotikov:1991pm, Remiddi:1997ny, Gehrmann:1999as, Argeri:2007up, MullerStach:2012mp, Henn:2013pwa, Henn:2014qga, Moriello:2019yhu, Hidding:2020ytt}, difference equations~\cite{Laporta:2000dsw, Lee:2009dh}, sector decomposition~\cite{Hepp:1966eg, Roth:1996pd, Binoth:2000ps, Heinrich:2008si, Smirnov:2015mct, Borowka:2015mxa, Borowka:2017idc} and Mellin-Barnes representation~\cite{Boos:1990rg, Smirnov:1999gc, Tausk:1999vh, Czakon:2005rk, Smirnov:2009up, Gluza:2007rt}, computation of MIs in these cutting-edge processes is still very challenging. For $\vec{s}$-DEs or difference equations method, one difficulty is that setting up these equations for very complicated processes becomes hard, and another difficulty is that there is no systematic way to obtain boundary conditions. While for the sector decomposition method or Mellin-Barnes representation method, numerical integration can be too inefficient for some processes and it is hard to get high precision.

In Ref.~\cite{Liu:2017jxz},  we proposed the auxiliary mass flow (AMF) method, a special case of the differential equations method. In the AMF method, MIs can be calculated by setting up and solving differential equations with respect to an auxiliary mass term $\eta$ (denoted as $\eta$-DEs) with almost trivial boundary conditions at $\eta\to\infty$. The method is systematic and efficient as far as $\eta$-DEs can be obtained. However, for complicated processes, the introduction of $\eta$ may greatly increase the number of MIs, such that the $\eta$-DEs cannot be set up in reasonable time with current reduction techniques~\cite{Chetyrkin:1981qh, Laporta:2000dsw, Gluza:2010ws, Schabinger:2011dz, vonManteuffel:2012np, Lee:2013mka, vonManteuffel:2014ixa, Larsen:2015ped, Peraro:2016wsq, Mastrolia:2018uzb, Liu:2018dmc,  Guan:2019bcx, Klappert:2019emp, Peraro:2019svx, Frellesvig:2019kgj, Wang:2019mnn, Smirnov:2019qkx, Klappert:2020nbg, Boehm:2020ijp, Heller:2021qkz, Bendle:2021ueg}. As a result, the application of this method is restricted to a large extent.

In this Letter, by introducing a key improvement to reduce the number of MIs, the AMF method becomes extremely powerful so that aforementioned cutting-edge problems are now solvable (see also Refs.~\cite{Song:2021vru, Chicherin:2018old,Papadopoulos:2015jft, Gehrmann:2018yef, Chicherin:2018mue, Chicherin:2018old, Chicherin:2020oor, Abreu:2020jxa, Canko:2020ylt} for recent developments on computations of MIs in these processes). Our method and results thus provide a valuable component for current and future high-precision phenomenological studies.

\sect{Auxiliary mass flow method with an iterative strategy}
Let us first give a brief review of the original AMF method proposed in Ref.~\cite{Liu:2017jxz}. For a given integral family, we introduce an auxiliary mass term to all of its propagators, which results in an auxiliary family of integrals
\begin{align}\label{eq:ampara}
{I}_{\text{aux}}(\vec{\nu};\eta)=\int\prod_{i=1}^{L}\frac{\ud^{D}\ell_i}{\ui\pi^{D/2}}
\frac{\mathcal{D}_{K+1}^{-\nu_{K+1}}\cdots \mathcal{D}_N^{-\nu_N}}{(\mathcal{D}_1- \eta)^{\nu_{1}}\cdots (\mathcal{D}_K- \eta)^{\nu_{K}}},
\end{align}
where $D = 4-2\epsilon$ is the spacetime dimension, $\mathcal{D}_1,\ldots,\mathcal{D}_K$ are inverse propagators of the original family, and $\mathcal{D}_{K+1},\ldots,\mathcal{D}_N$ are irreducible scalar products introduced for completeness. Physical integrals are defined by
\begin{align}\label{eq:def}
{I}(\vec{\nu})= \lim_{\eta\to \ui 0^-}{I}_{\text{aux}}(\vec{\nu};\eta).
\end{align}
The introduction of $\eta$ brings a nice feature: when $|\eta|$ is large enough, all integrals in the auxiliary family can be expanded as linear combinations of equal-mass vacuum integrals, which have been well studied \cite{Davydychev:1992mt, Broadhurst:1998rz, Schroder:2005va, Luthe:2015ngq, Kniehl:2017ikj, Luthe:2017ttc}. Then physical MIs can be obtained from MIs of the auxiliary family, denoted as $\vec{\mathcal{I}}_\text{aux}(\eta)$, by the flow of $\eta$ from $\infty$ to $\ui0^-$, which can be realized by setting up and solving a system of $\eta$-DEs,
\begin{align}\label{eq:deq}
 \frac{\partial}{\partial \eta}\vec{\mathcal{I}}_\text{aux}(\eta) = A(\eta)\vec{\mathcal{I}}_\text{aux}(\eta).
\end{align}
Unfortunately, the introduction of $\eta$ usually significantly increases the number of MIs, which makes it hard to set up the above $\eta$-DEs for very complicated processes.

\begin{figure}[htbp]
\centering
\includegraphics[width = 0.25\textwidth]{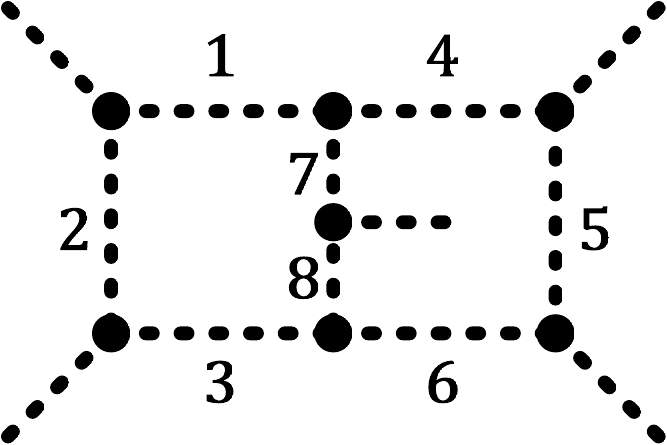}\\
\caption{Massless two-loop double-pentagon Feynman diagram.}\label{fig:dp}
\end{figure}

Our improvement is based on a simple observation that the number of MIs can be reduced if we introduce $\eta$ to fewer propagators. Taking the massless two-loop double-pentagon family in Fig.~\ref{fig:dp} as an example, which contains totally 108 MIs before introducing $\eta$, we summarize relevant information  in Tab.~\ref{tab:dp} for introducing $\eta$ in several different ways. Here, the four different modes, ``all'', ``loop'', ``branch'' or ``propagator'' correspond  to introducing $\eta$ to all propagators, propagators forming a closed loop, propagators on the same branch, or a single propagator, respectively. It is clear that the newly introduced modes can indeed result in fewer MIs than the original ``all'' mode. Furthermore, we find that the computational cost to construct and solve the system of $\eta$-DEs is usually positively correlated to the number of MIs, which is as expected. As a result, new modes can provide great improvement to the AMF method.

We note that, besides the above modes, sometimes a ``mass'' mode can be defined, which  adds $\eta$ to propagators with equal nonzero mass, with the condition that this mass is independent of scales from external momenta. As the ``mass'' mode does not increase the number of MIs, it should be used as far as possible, e.g., Refs.~\cite{Bronnum-Hansen:2020mzk, Bronnum-Hansen:2021olh}. But for the current example, there is no massive propagator to enable the use of the ``mass'' mode.

\begin{table}[htbp]
	\centering
	\begin{tabular}{|c|c|c|}\hline
    Mode & Propagators         & No. of MIs\\\hline
    all  & \{1,2,3,4,5,6,7,8\} & 476\\\hline
    \multirow{2}{*}{loop} & \{4,5,6,7,8\} & 305\\\cline{2-3}
                          & \{1,2,3,4,5,6\} & 319\\\hline
    \multirow{2}{*}{branch} & \{4,5,6\} & 233\\\cline{2-3}
                          & \{7,8\} & 234\\\hline

    \multirow{3}{*}{propagator} & \{4\} & 178\\\cline{2-3}
                          & \{5\} & 176\\\cline{2-3}
                          & \{7\} & 220\\\hline

    \multirow{1}{*}{mass} & - & -\\\hline

	\end{tabular}
	\caption{Number of MIs of double-pentagon family after introducing $\eta$ in different ways. The MIs are determined through \textsf{LiteRed}~\cite{Lee:2013mka} and \textsf{FiniteFlow}~\cite{Peraro:2019svx}.}
	\label{tab:dp}
\end{table}

In a general mode, however, MIs of the auxiliary family cannot be simply decomposed to linear combinations of vacuum integrals near $\eta = \infty$, because there are usually more integration regions. Fortunately,  when $|\eta|$ is very large, all integration regions can be systematically identified, following the general rules of region analysis~\cite{Beneke:1997zp, Smirnov:1999bza}. Actually, inequivalent regions can be characterized by the size of loop momentum carried by each branch of the diagram, which can be either of $\mathcal{O}(\sqrt{\eta})$~(denoted as large loop momentum, L) or $\mathcal{O}(1)$~(denoted as small loop momentum, S). For example, in one-loop case, there is only one branch with loop momentum $\ell_1$. As a result, two integration regions may contribute: (L) and (S). In two-loop case, there are in general three branches, whose loop momenta are $\ell_1$, $\ell_2$ and $\ell_1+\ell_2$ respectively. So, five regions at most may contribute: (LLL), (LLS), (LSL), (SLL) and (SSS), in consideration of the fact that (LSS), (SLS) and (SSL) are excluded by momentum conservation. Regions in higher-loop cases can also be figured out analogously.

To obtain boundary conditions near $\eta = \infty$, we expand integrands in each region. Specifically, in the all-large region (L...L), each propagator is expanded as
\begin{align}\label{eq:alllarge}
\frac{1}{(\ell+p)^2-m^2- \kappa\,\eta }\sim \frac{1}{\ell^2- \kappa\,\eta },
\end{align}
where $\ell$ is a linear combination of loop momenta, and $\kappa = 0$ or $1$, depending on whether $\eta$ is introduced to this propagator. We thus obtain vacuum integrals after expansion in this region.
In the all-small region (S...S), only propagators containing $\eta$ should be expanded as
\begin{align}\label{eq:small}
\frac{1}{(\ell+p)^2-m^2-\eta}\sim \frac{1}{-\eta}.
\end{align}
In this case, we get integrals in a subfamily.
In mixed regions, we decompose loop momentum of each propagator as the sum of a large part $\ell_\text{L}$ and a small part $\ell_\text{S}$. Then, if $\ell_\text{L}\neq 0$ or $\kappa\neq 0$, we can expand the propagator as
\begin{align}\label{eq:mixed1}
\frac{1}{(\ell_\text{L}+\ell_\text{S}+p)^2-m^2- \kappa\,\eta}&\sim \frac{1}{\ell_\text{L}^2- \kappa\,\eta}.
\end{align}
Otherwise, no expansion is needed. So, after the expansion in mixed regions, the part containing large loop momenta and the part containing small loop momenta are decoupled, which results in factorized integrals.
It is now clear that region analysis for the ``all'' mode is just a special case, where only the region (L...L) contributes.

\begin{figure}[htbp]
\centering
\includegraphics[width=0.45\textwidth]{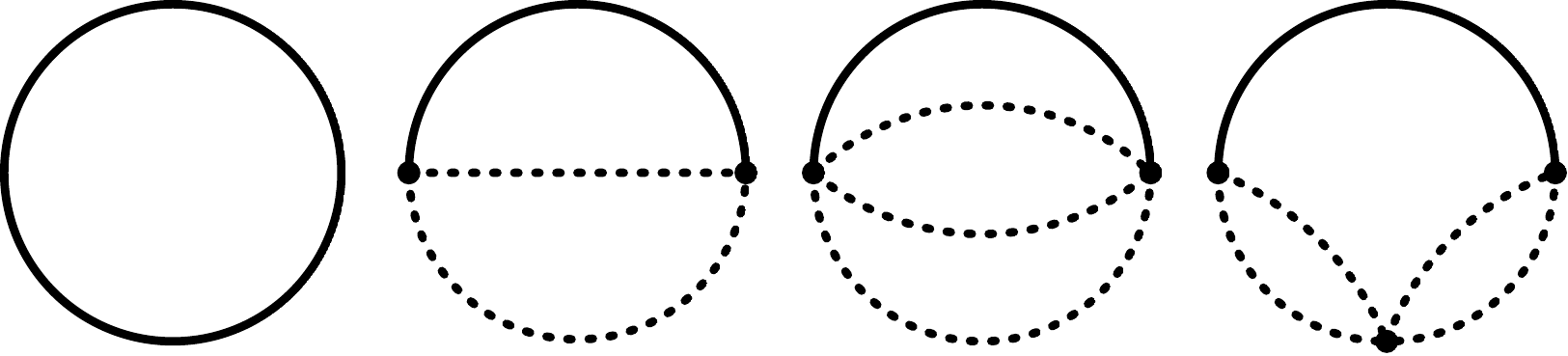}\\
\caption{Nonfactorized MIs for single-mass vacuum integrals up to three loops. The solid curves represent massive propagators and dashed curves represent massless propagators.}\label{fig:singlevac}
\end{figure}

Although the obtained integrals after expansion are more complicated than those in the ``all'' mode, especially the integrals obtained in all-small region, it does not bring any trouble because they have already been simpler than original integrals. Then we can further introduce $\eta$ to these obtained integrals and use $\eta$-DEs to push boundary conditions to even simpler integrals. After several steps of iteration, we can always cast boundary integrals to either scaleless integrals or single-scale vacuum integrals. The former ones can be set to zero in dimensional regularization directly, and the latter ones are simple enough to calculate. Actually, we can always use the ``propagator'' mode to cast boundary conditions to single-mass vacuum integrals, as shown in Fig.~\ref{fig:singlevac}, which can be related to extensively studied p-integrals \cite{Georgoudis:2021onj}.

Eventually, we can numerically solve the systems of $\eta$-DEs, iteratively, to obtain physical MIs. We emphasize that the step to take $\eta\to\ui0^-$ in any mode is similar to that for the ``all mode'' discussed in details in Ref.~\cite{Liu:2017jxz}. Using $\eta$-DEs, we can get generalized power series expansion at $\eta=0$, and then, to obtain the limit $\eta\to\ui0^-$, we can safely discard all nonanalytical terms in $\eta$ in the spirit of dimensional regularization.

\sect{A pedagogical example}
As a pedagogical example, we show how to calculate MIs in the massless two-loop double-pentagon family, as shown in Fig.~\ref{fig:dp}, using AMF method with the new iterative strategy. Note that MIs of this family were successfully calculated only in the last two years~\cite{Chicherin:2018old} taking advantage of the powerful canonical form of DEs~\cite{Henn:2013pwa}. The family contains five external momenta $\{p_1,\ldots,p_5\}$ flowing outside, which satisfy the on-shell conditions $p_i^2 = 0$ and momentum conservation $\sum_{i=1}^5 p_i = 0$. Thus there are five independent kinematic variables $\vec{s} = \{s_{12}, s_{23}, s_{34}, s_{45}, s_{15}\}$ with $s_{ij} =  (p_i+p_j)^2$. We denote
\begin{align}
&\mathcal{D}_1 = \ell_1^2, \,\mathcal{D}_2 = (\ell_1-p_1)^2,\, \mathcal{D}_3 = (\ell_1-p_1-p_2)^2, \nonumber\\
&\mathcal{D}_4 = \ell_2^2,\, \mathcal{D}_5 = (\ell_2+p_5)^2,\, \mathcal{D}_6 = (\ell_2+p_4+p_5)^2,\nonumber\\
&\mathcal{D}_7 = (\ell_1-\ell_2)^2,\,\mathcal{D}_8 = (\ell_1-\ell_2+p_3)^2,\,\mathcal{D}_9 = (\ell_1+p_5)^2,\nonumber\\
&\mathcal{D}_{10} = (\ell_2-p_1)^2,\,\mathcal{D}_{11} = (\ell_2-p_1-p_2)^2,
\end{align}
where the first eight are inverse propagators and the last three are irreducible scalar products.

As shown in Tab.~\ref{tab:dp}, it should be a good choice to apply the ``propagator'' mode to first introduce $\eta$ to $\mathcal{D}_5$. Then by expanding integrands of MIs of the auxiliary family in each integration region, we find only (LLL), (SLL) and (SSS) contribute, because all other regions are scaleless. In the region (LLL), all obtained integrals after expansion can be reduced to the two-loop single-mass vacuum MI in Fig.~\ref{fig:singlevac} directly. While in the region (SLL), after integral reduction, only a factorized integral remains
\begin{align}\label{eq:SLLbi}
\int \frac{\ud^D\ell_1}{\ui\pi^{D/2}}\frac{1}{\ell_1^2(\ell_1-p_1-p_2)^2} \times \int\frac{\ud^D\ell_2}{\ui\pi^{D/2}}\frac{1}{\ell_2^2-1}.
\end{align}
Although this integral is simple enough to calculate, we can again introduce $\eta$ to its nonvacuum part to cast it to a vacuum-integral problem. This step is quite useful in practice to make the computations more automatic.

\begin{figure}[htbp]
\centering
\includegraphics[width=0.45\textwidth]{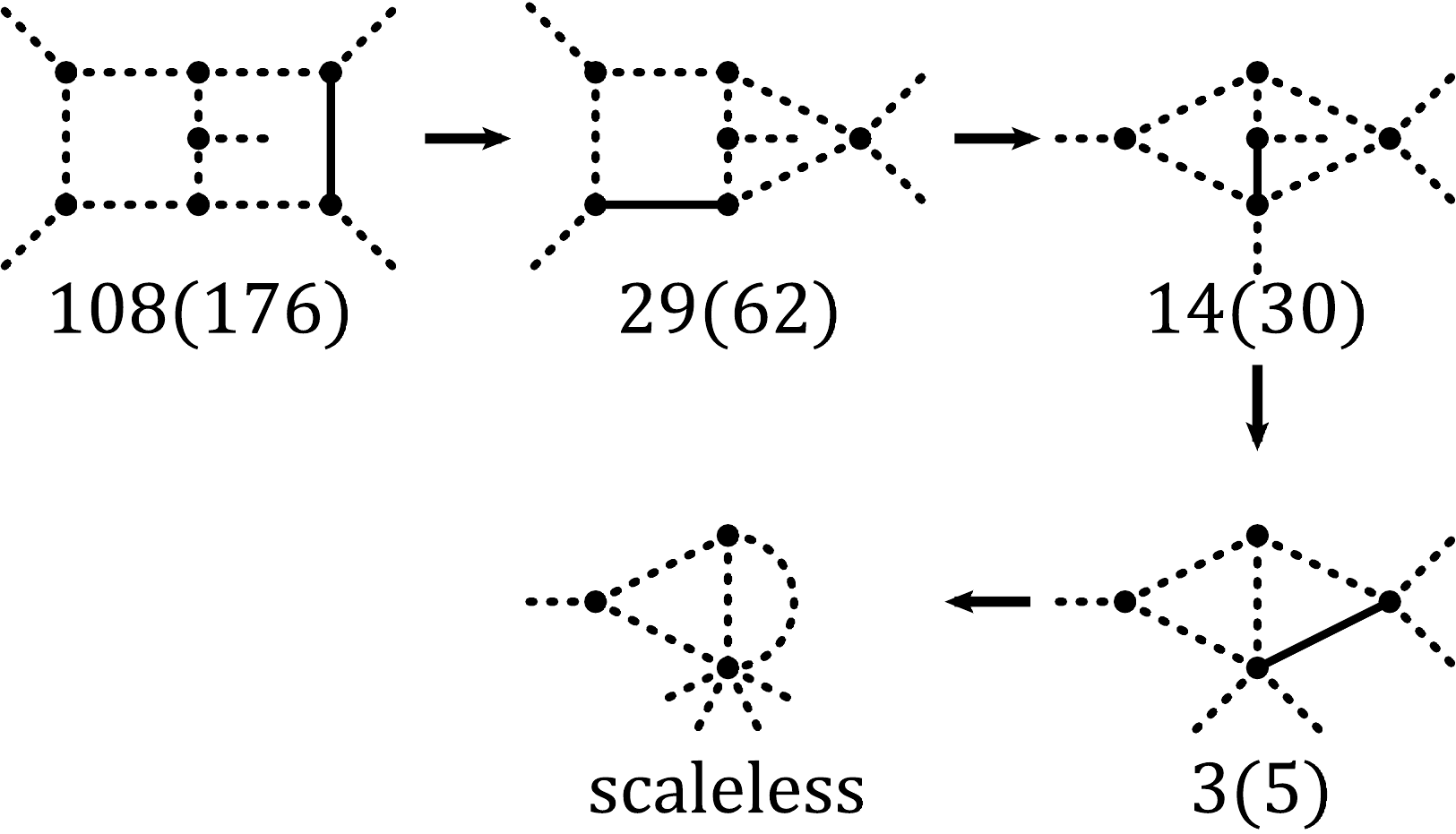}\\
\caption{The all-small region iteration of massless two-loop double-pentagon in ``propagator'' mode. Solid lines mean propagators with $\eta$. The number of master integrals before and after introducing $\eta$ are shown under each diagram, except the last one.}\label{fig:dpSiter}
\end{figure}

In the most complicated region (SSS), we need to handle integrals in the subfamily obtained by contracting the fifth propagator. To calculate these integrals, we again apply the ``propagator'' mode to further introduce $\eta$ to $\mathcal{D}_3$. The region analysis is the same as before. The only nontrivial one is still the (SSS) region, where a six-propagator subfamily emerges. Then, we can repeat the above steps to simplify the expanded integrals in the (SSS) region iteratively. We show this iteration in Fig.~\ref{fig:dpSiter}, where solid lines represent propagators with $\eta$ introduced. We also present the number of original MIs along with that after introducing $\eta$ under each diagram. The last step is a scaleless subfamily. So, after this iteration, all integrals after expansion around $\eta\to \infty$ are known to us.

To obtain a numerical result, we choose a physical phase-space point at
\begin{align}
\vec{s}_0 = \left\{4, -\frac{113}{47}, \frac{281}{149}, \frac{349}{257}, -\frac{863}{541} \right\}.
\end{align}
The systems of $\eta$-DEs are constructed with this numerical configuration for simplicity, i.e., they are only analytic in $\eta$ and space-time dimension $D$. Using our package \texttt{AMFlow} \cite{Liu:2022chg}, we numerically solve the systems of $\eta$-DEs to obtain the physical results. The following is the result of a top sector integral truncated at $\mathcal{O}(\epsilon^5)$ with 16-digit precision,
\begin{align}\label{eq:numerical}
& I(1,1,1,1,1,1,1,1,0,0,0) =\nonumber\\
& -0.06943562517263776\epsilon^{-4}\nonumber\\
& +(1.162256636711287+1.416359853446717 \ui)\epsilon^{-3}\nonumber\\
& +(37.82474332116938+15.91912443581739 \ui)\epsilon^{-2}\nonumber\\
& +(86.2861798369034+166.8971535711277 \ui)\epsilon^{-1}\nonumber\\
& -(4.1435965578662-333.0996040071305 \ui)\nonumber\\
& -(531.834114822928-1583.724672502141 \ui)\epsilon\nonumber\\
& -(2482.240253232612-2567.398291724192 \ui)\epsilon^2\nonumber\\
& -(8999.90369367113-19313.42643829926 \ui) \epsilon^3\nonumber\\
& -(28906.95582696762-17366.82954322838 \ui) \epsilon^4,
\end{align}
while results of other integrals can be found in the ancillary file. We note that results with higher precision and with higher order in $\mathcal{\epsilon}$ expansion can be achieved easily within our method.

The coefficients of $\mathcal{\epsilon}^{n}$ with $n\leq 0$ in Eq.~\eqref{eq:numerical} are found to agree with analytic results obtained in Ref.~\cite{Chicherin:2018old}.
The coefficients of $\mathcal{\epsilon}^{n}$ with $n> 0$ are new, and they have passed a highly nontrivial self-consistency check. On the one hand, results at another phase space point $\vec{s}_1$ can be obtained by solving a system of one-dimension $\vec{s}$-DEs~\cite{Moriello:2019yhu,Bonciani:2019jyb,Frellesvig:2019byn,Abreu:2020jxa} along the line between $\vec{s}_0$ and $\vec{s}_1$, with boundary conditions at $\vec{s}_0$. On the other hand, results at $\vec{s}_1$ can also be calculated directly using the AMF method. We find full agreement between results obtained in the two ways.

\begin{figure}[htbp]
\centering
\includegraphics[width=0.45\textwidth]{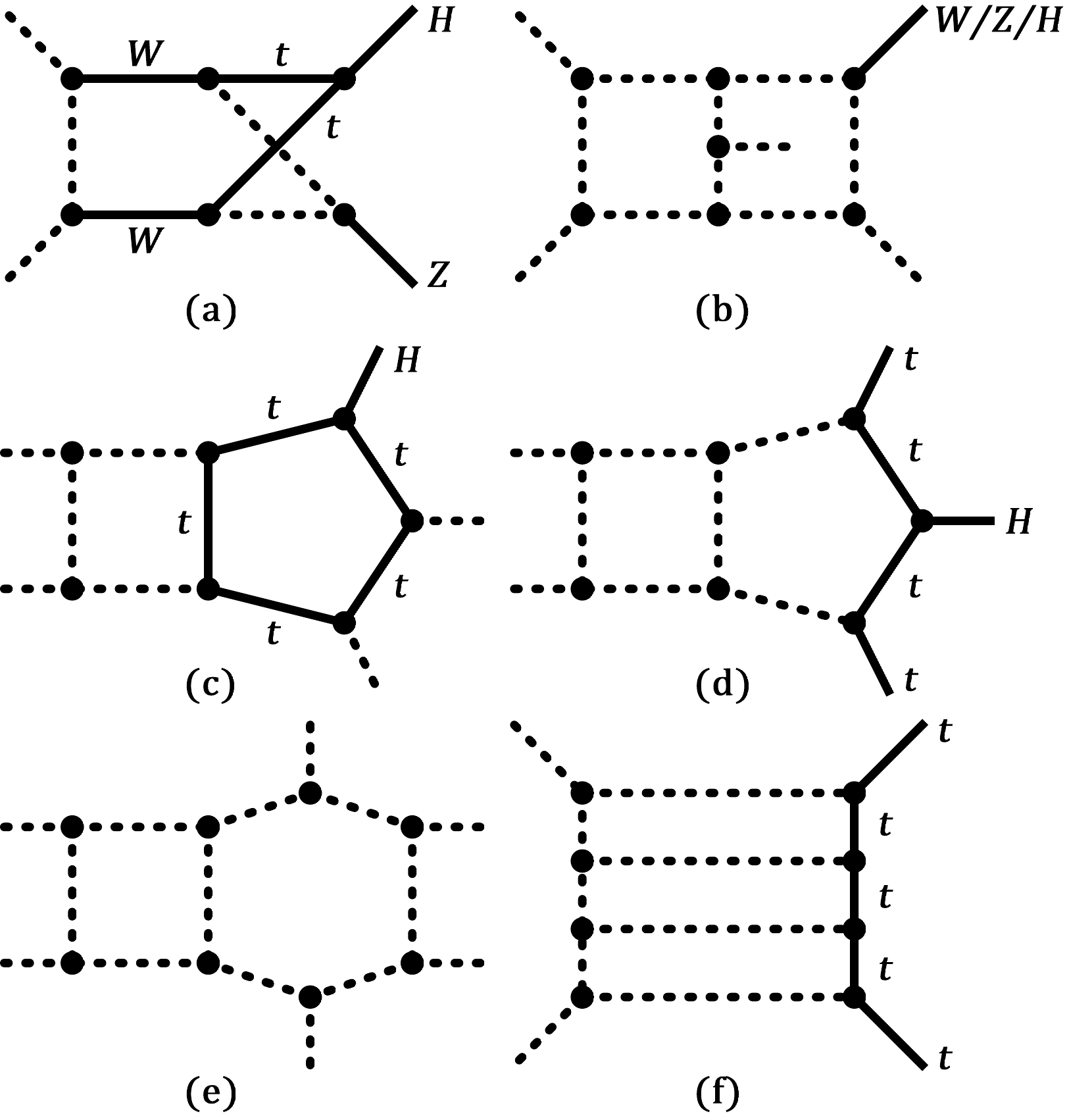}\\
\caption{Some topologies involved in cutting-edge processes. Massless particles and massive particles are denoted as dashed lines and solid lines, respectively. }\label{fig:examples}
\end{figure}

\sect{Applications on cutting-edge problems}
Now, we use the improved AMF method to calculate MIs encountered in some cutting-edge problems.
As shown in Fig.~\ref{fig:examples}, integral family (a) is relevant for two-loop electroweak corrections to $e^+e^-\to HZ$,  families (b)-(e) are relevant for two-loop QCD corrections to $W/Z/H+2j$,  $H+2j$,$t\bar{t}H$ and $4j$ production at hadron colliders, respectively; and family (f) is relevant for three-loop QCD corrections to $t\bar{t}$ production at hadron colliders. All these processes are very important in the following decade. Among them, MIs of the family (a) could in principle be calculated using the method developed in Ref.~\cite{Song:2021vru}; while MIs of the rest topologies may be challenging for all methods on the market and have not been calculated yet as far as we know.

To calculate MIs in families (a) and (c), we choose the ``mass'' mode at the first step of the iteration, because there are masses in denominators that are independent of external scales. Other families are calculated using the ``propagator'' mode. All $\eta$-DEs are constructed with arbitrarily chosen numerical external kinematics and MIs are evaluated with 16-digit precision with $\epsilon$ expanded up to $\epsilon^4$. The results have passed the self-consistency check, as explained in details for the double-pentagon family. Some subsector integrals, e.g., in topology (b), are also checked with results in the literature~\cite{Abreu:2020jxa, Canko:2020ylt}. Our results are provided in the supplemental material.

On the one hand, our results can serve as high-precision boundary conditions for $\vec{s}$-DEs if fully analytic $\vec{s}$-DEs can be constructed, and thus MIs at any given phase space points can be achieved. On the other hand, if $\vec{s}$-DEs are hard to obtain, MIs at any given phase space point can be calculated directly using the  AMF method. Therefore, our results are important building blocks and benchmarks for various future studies.

A few comments are in order:

1) Time consumptions for setting up and solving all $\eta$-DEs are summarized in Tab.~\ref{tab:all},  along with percentages at the first step in the iteration. It is clear that the first step dominates the cost in any family, and the cost of all later steps are negligible. This phenomenon implies that the ``propagator'' mode is usually the best choice, if a ``mass'' mode does not exist, because it may results in the minimal number of MIs at the first step.

\begin{table}[htbp]
	\centering
	\begin{tabular}{|c|c|c|c|c|c|c|c|}\hline
		Family         & dp   & (a) & (b) & (c) & (d) & (e) & (f) \\\hline
		$T_\text{setup}$ & 6 & 20  & 18  &  8  &  1  &  25 &  30 \\\hline
		$T_\text{solve}$ & 7 & 11  & 15  &  6  &  3  &  15 &  42 \\\hline
		$P_1$             &95\%  & 99\%& 96\%& 99\%& 98\%& 94\%& 93\%\\\hline\hline
        $T_{\vec{s}}$   &  2  &  916  & 64  &  1305 &  30  & 1801   & 63  \\\hline
	\end{tabular}
	\caption{Time consumption to compute various families, where ``dp'' denotes the double-pentagon family in Fig.~\ref{fig:dp} and ``(a)''-``(f)'' denote corresponding families in Fig.~\ref{fig:examples}. $T_\text{setup}$ denotes time consumption to set up $\eta$-DEs, including steps to construct the block-triangular linear systems, $T_\text{solve}$ denotes time consumption to numerically solve the $\eta$-DEs to obtain 16-digit precision results, $P_1$ denotes the percentage to set up and solve $\eta$-DEs at the first step in the iteration, and $T_{\vec{s}}$ denotes time consumption to set up one-dimension $\vec{s}$-DEs along the line between two chosen phase space points. Time consumption  is counted in the unit of CPU core hours.}
	\label{tab:all}
\end{table}

2) Differential equations are solved numerically using our  \textsf{Mathematica} package \texttt{AMFlow}~\cite{Liu:2022chg}. We find the number of correct digits is in proportion to the time consumption, due to the fact that $\eta$-DEs are rational functions of $\eta$.
Furthermore, it can be expected that the efficiency can be significantly improved if a low-level language, like \textsf{Fortran}, is used. Therefore, the number of digits is rather cheap in the AMF method, and reconstruction of analytic results via the PSLQ algorithm (see Ref.~\cite{Bailey:1999nv} and references therein) might be possible for problems with not too many mass scales, which will be investigated in the future.

3) Integrals reduction is needed to set up DEs. We utilize the method developed in Refs.~\cite{Liu:2018dmc, Guan:2019bcx} to construct block-triangular linear systems among Feynman integrals, which are found to be much more efficient than traditional integration-by-parts systems constructed by \textsf{LiteRed} \cite{Lee:2013mka} and solved by \textsf{FiniteFlow}\cite{Peraro:2019svx}. To set up $\eta$-DEs in this Letter, the block-triangular systems can roughly speed up by 2 orders of magnitude.

4) Time consumptions for constructing systems of one-dimension $\vec{s}$-DEs of physical MIs are also summarized in Tab.~\ref{tab:all}. It looks first surprising that they are usually larger than the costs to set up $\eta$-DEs, although $\eta$-DEs may involve more MIs. This is understandable because the derivative of Feynman integrals with respect to $\eta$ only increase the degree of denominators by 1, but that with respect to $\vec{s}$ increases not only the degree of denominators but also the degree of numerators, which results in a much larger reduction system that is  much harder to solve.

\sect{Summary and outlook}
We find that, by introducing the auxiliary mass $\eta$ to fewer propagators at a time, the AMF method has a much better performance, and meanwhile the systematics of AMF method is untouched. We show that the AMF method can provide high-precision boundary conditions for traditional DEs as well as provide a highly nontrivial self-consistency check. Furthermore, an interesting observation is that it is usually easier to set up $\eta$-DEs comparing with $\vec{s}$-DEs. Therefore, the AMF method may still work even if a problem is so difficult that $\vec{s}$-DEs becomes hard to set up.

Equipped with the improved AMF method, we have successfully calculated some very difficult Feynman integrals encountered in two-loop electroweak corrections to $e^+e^-\to HZ$, two-loop QCD corrections to $3j$, $W/Z/H+2j$, $t\bar{t}H$ and $4j$ production at hadron colliders, and three-loop QCD corrections to $t\bar{t}$ production at hadron colliders. As all these processes are crucial in the following decade, our results are important building blocks and benchmarks for future studies.

Our method has been implemented in our package \texttt{AMFlow}~\cite{Liu:2022chg}.
In the near future, with visible developments of integral reduction techniques, we expect that master integrals of all next-generation processes can be calculable using the AMF method.

\sect{Acknowledgments}
We thank K.T. Chao, Z. Li, X.H. Liu, Z.F. Liu, C. Meng, W. Tao, R.H. Wu, G. Yang, P. Zhang and Y. Zhang for many useful communications and discussions. The work is supported in part by the National Natural Science Foundation of China (Grants No. 11875071 and No. 11975029), the National Key Research and Development Program of China under Contracts No. 2020YFA0406400, and the High-performance Computing Platform of Peking University.

\providecommand{\href}[2]{#2}\begingroup\raggedright\endgroup


\begin{thebibliography}{10}

\bibitem{Cepeda:2019klc}
M.~Cepeda {\em et al.}, {\it {Report from Working Group 2}: {Higgs Physics at
  the HL-LHC and HE-LHC}},
  \href{http://dx.doi.org/10.23731/CYRM-2019-007.221}{{\em CERN Yellow Rep.
  Monogr.} {\bfseries 7} (2019) 221--584}
  [\href{http://arxiv.org/abs/1902.00134}{{\ttfamily arXiv:1902.00134}}]
  [\href{http://inspirehep.net/search?p=find+Cepeda:2019klc}{{\ttfamily
  InSPIRE}}].

\bibitem{CEPCStudyGroup:2018ghi}
{\bfseries CEPC Study Group} , M.~Dong {\em et al.}, {\it {CEPC Conceptual
  Design Report: Volume 2 - Physics \& Detector}},
  [\href{http://arxiv.org/abs/1811.10545}{{\ttfamily arXiv:1811.10545}}]
  [\href{http://inspirehep.net/search?p=find+CEPCStudyGroup:2018ghi}{{\ttfamily
  InSPIRE}}].

\bibitem{Kotikov:1990kg}
A.~V. Kotikov, {\it {Differential equations method: New technique for massive
  Feynman diagrams calculation}},
\href{http://dx.doi.org/10.1016/0370-2693(91)90413-K}{{\em Phys. Lett.}
  {\bfseries B254} (1991) 158--164}
  [\href{http://inspirehep.net/search?p=find+Kotikov:1990kg}{{\ttfamily
  InSPIRE}}].

\bibitem{Kotikov:1991pm}
A.~V. Kotikov, {\it {Differential equation method: The Calculation of N point
  Feynman diagrams}},
  \href{http://dx.doi.org/10.1016/0370-2693(91)90536-Y}{{\em Phys. Lett. B}
  {\bfseries 267} (1991) 123--127}
  [\href{http://inspirehep.net/search?p=find+Kotikov:1991pm}{{\ttfamily
  InSPIRE}}]. [Erratum: Phys.Lett.B 295, 409--409 (1992)].

\bibitem{Remiddi:1997ny}
E.~Remiddi, {\it {Differential equations for Feynman graph amplitudes}},
{\em Nuovo Cim.} {\bfseries A110} (1997) 1435--1452
  [\href{http://arxiv.org/abs/hep-th/9711188}{{\ttfamily hep-th/9711188}}]
  [\href{http://inspirehep.net/search?p=find+Remiddi:1997ny}{{\ttfamily
  InSPIRE}}].

\bibitem{Gehrmann:1999as}
T.~Gehrmann and E.~Remiddi, {\it {Differential equations for two loop four
  point functions}},
\href{http://dx.doi.org/10.1016/S0550-3213(00)00223-6}{{\em Nucl. Phys.}
  {\bfseries B580} (2000) 485--518}
  [\href{http://arxiv.org/abs/hep-ph/9912329}{{\ttfamily hep-ph/9912329}}]
  [\href{http://inspirehep.net/search?p=find+Gehrmann:1999as}{{\ttfamily
  InSPIRE}}].

\bibitem{Argeri:2007up}
M.~Argeri and P.~Mastrolia, {\it {Feynman Diagrams and Differential
  Equations}},  \href{http://dx.doi.org/10.1142/S0217751X07037147}{{\em Int. J.
  Mod. Phys. A} {\bfseries 22} (2007) 4375--4436}
  [\href{http://arxiv.org/abs/0707.4037}{{\ttfamily arXiv:0707.4037}}]
  [\href{http://inspirehep.net/search?p=find+Argeri:2007up}{{\ttfamily
  InSPIRE}}].

\bibitem{MullerStach:2012mp}
S.~M\"uller-Stach, S.~Weinzierl, and R.~Zayadeh, {\it {Picard-Fuchs equations
  for Feynman integrals}},
\href{http://dx.doi.org/10.1007/s00220-013-1838-3}{{\em Commun. Math. Phys.}
  {\bfseries 326} (2014) 237--249}
  [\href{http://arxiv.org/abs/1212.4389}{{\ttfamily arXiv:1212.4389}}]
  [\href{http://inspirehep.net/search?p=find+MullerStach:2012mp}{{\ttfamily
  InSPIRE}}].

\bibitem{Henn:2013pwa}
J.~M. Henn, {\it {Multiloop integrals in dimensional regularization made
  simple}},
\href{http://dx.doi.org/10.1103/PhysRevLett.110.251601}{{\em Phys. Rev. Lett.}
  {\bfseries 110} (2013) 251601}
  [\href{http://arxiv.org/abs/1304.1806}{{\ttfamily arXiv:1304.1806}}]
  [\href{http://inspirehep.net/search?p=find+Henn:2013pwa}{{\ttfamily
  InSPIRE}}].

\bibitem{Henn:2014qga}
J.~M. Henn, {\it {Lectures on differential equations for Feynman integrals}},
\href{http://dx.doi.org/10.1088/1751-8113/48/15/153001}{{\em J. Phys.}
  {\bfseries A48} (2015) 153001}
  [\href{http://arxiv.org/abs/1412.2296}{{\ttfamily arXiv:1412.2296}}]
  [\href{http://inspirehep.net/search?p=find+Henn:2014qga}{{\ttfamily
  InSPIRE}}].

\bibitem{Moriello:2019yhu}
F.~Moriello, {\it {Generalised power series expansions for the elliptic planar
  families of Higgs + jet production at two loops}},
  \href{http://dx.doi.org/10.1007/JHEP01(2020)150}{{\em JHEP} {\bfseries 01}
  (2020) 150} [\href{http://arxiv.org/abs/1907.13234}{{\ttfamily
  arXiv:1907.13234}}]
  [\href{http://inspirehep.net/search?p=find+Moriello:2019yhu}{{\ttfamily
  InSPIRE}}].

\bibitem{Hidding:2020ytt}
M.~Hidding, {\it {DiffExp, a Mathematica package for computing Feynman
  integrals in terms of one-dimensional series expansions}},
  \href{http://dx.doi.org/10.1016/j.cpc.2021.108125}{{\em Comput. Phys.
  Commun.} {\bfseries 269} (2021) 108125}
  [\href{http://arxiv.org/abs/2006.05510}{{\ttfamily arXiv:2006.05510}}]
  [\href{http://inspirehep.net/search?p=find+Hidding:2020ytt}{{\ttfamily
  InSPIRE}}].

\bibitem{Laporta:2000dsw}
S.~Laporta, {\it {High precision calculation of multiloop Feynman integrals by
  difference equations}},
  \href{http://dx.doi.org/10.1142/S0217751X00002159}{{\em Int. J. Mod. Phys. A}
  {\bfseries 15} (2000) 5087--5159}
  [\href{http://arxiv.org/abs/hep-ph/0102033}{{\ttfamily hep-ph/0102033}}]
  [\href{http://inspirehep.net/search?p=find+Laporta:2000dsw}{{\ttfamily
  InSPIRE}}].

\bibitem{Lee:2009dh}
R.~N. Lee, {\it {Space-time dimensionality D as complex variable: Calculating
  loop integrals using dimensional recurrence relation and analytical
  properties with respect to D}},
\href{http://dx.doi.org/10.1016/j.nuclphysb.2009.12.025}{{\em Nucl. Phys.}
  {\bfseries B830} (2010) 474--492}
  [\href{http://arxiv.org/abs/0911.0252}{{\ttfamily arXiv:0911.0252}}]
  [\href{http://inspirehep.net/search?p=find+Lee:2009dh}{{\ttfamily InSPIRE}}].

\bibitem{Hepp:1966eg}
K.~Hepp, {\it {Proof of the Bogolyubov-Parasiuk theorem on renormalization}},
  \href{http://dx.doi.org/10.1007/BF01773358}{{\em Commun. Math. Phys.}
  {\bfseries 2} (1966) 301--326}
  [\href{http://inspirehep.net/search?p=find+Hepp:1966eg}{{\ttfamily
  InSPIRE}}].

\bibitem{Roth:1996pd}
M.~Roth and A.~Denner, {\it {High-energy approximation of one loop Feynman
  integrals}},  \href{http://dx.doi.org/10.1016/0550-3213(96)00435-X}{{\em
  Nucl. Phys. B} {\bfseries 479} (1996) 495--514}
  [\href{http://arxiv.org/abs/hep-ph/9605420}{{\ttfamily hep-ph/9605420}}]
  [\href{http://inspirehep.net/search?p=find+Roth:1996pd}{{\ttfamily
  InSPIRE}}].

\bibitem{Binoth:2000ps}
T.~Binoth and G.~Heinrich, {\it {An automatized algorithm to compute infrared
  divergent multiloop integrals}},
\href{http://dx.doi.org/10.1016/S0550-3213(00)00429-6}{{\em Nucl. Phys.}
  {\bfseries B585} (2000) 741--759}
  [\href{http://arxiv.org/abs/hep-ph/0004013}{{\ttfamily hep-ph/0004013}}]
  [\href{http://inspirehep.net/search?p=find+Binoth:2000ps}{{\ttfamily
  InSPIRE}}].

\bibitem{Heinrich:2008si}
G.~Heinrich, {\it {Sector Decomposition}},
\href{http://dx.doi.org/10.1142/S0217751X08040263}{{\em Int. J. Mod. Phys.}
  {\bfseries A23} (2008) 1457--1486}
  [\href{http://arxiv.org/abs/0803.4177}{{\ttfamily arXiv:0803.4177}}]
  [\href{http://inspirehep.net/search?p=find+Heinrich:2008si}{{\ttfamily
  InSPIRE}}].

\bibitem{Smirnov:2015mct}
A.~V. Smirnov, {\it {FIESTA4: Optimized Feynman integral calculations with GPU
  support}},
\href{http://dx.doi.org/10.1016/j.cpc.2016.03.013}{{\em Comput. Phys. Commun.}
  {\bfseries 204} (2016) 189--199}
  [\href{http://arxiv.org/abs/1511.03614}{{\ttfamily arXiv:1511.03614}}]
  [\href{http://inspirehep.net/search?p=find+Smirnov:2015mct}{{\ttfamily
  InSPIRE}}].

\bibitem{Borowka:2015mxa}
S.~Borowka, G.~Heinrich, S.~P. Jones, M.~Kerner, J.~Schlenk, and T.~Zirke, {\it
  {SecDec-3.0: numerical evaluation of multi-scale integrals beyond one loop}},
\href{http://dx.doi.org/10.1016/j.cpc.2015.05.022}{{\em Comput. Phys. Commun.}
  {\bfseries 196} (2015) 470--491}
  [\href{http://arxiv.org/abs/1502.06595}{{\ttfamily arXiv:1502.06595}}]
  [\href{http://inspirehep.net/search?p=find+Borowka:2015mxa}{{\ttfamily
  InSPIRE}}].

\bibitem{Borowka:2017idc}
S.~Borowka, G.~Heinrich, S.~Jahn, S.~P. Jones, M.~Kerner, J.~Schlenk, and
  T.~Zirke, {\it {pySecDec: a toolbox for the numerical evaluation of
  multi-scale integrals}},
\href{http://dx.doi.org/10.1016/j.cpc.2017.09.015}{{\em Comput. Phys. Commun.}
  {\bfseries 222} (2018) 313--326}
  [\href{http://arxiv.org/abs/1703.09692}{{\ttfamily arXiv:1703.09692}}]
  [\href{http://inspirehep.net/search?p=find+Borowka:2017idc}{{\ttfamily
  InSPIRE}}].

\bibitem{Boos:1990rg}
E.~E. Boos and A.~I. Davydychev, {\it {A Method of evaluating massive Feynman
  integrals}},  \href{http://dx.doi.org/10.1007/BF01016805}{{\em Theor. Math.
  Phys.} {\bfseries 89} (1991) 1052--1063}
  [\href{http://inspirehep.net/search?p=find+Boos:1990rg}{{\ttfamily
  InSPIRE}}].
[Teor. Mat. Fiz.89,56(1991)].

\bibitem{Smirnov:1999gc}
V.~A. Smirnov, {\it {Analytical result for dimensionally regularized massless
  on shell double box}},
\href{http://dx.doi.org/10.1016/S0370-2693(99)00777-7}{{\em Phys. Lett.}
  {\bfseries B460} (1999) 397--404}
  [\href{http://arxiv.org/abs/hep-ph/9905323}{{\ttfamily hep-ph/9905323}}]
  [\href{http://inspirehep.net/search?p=find+Smirnov:1999gc}{{\ttfamily
  InSPIRE}}].

\bibitem{Tausk:1999vh}
J.~B. Tausk, {\it {Nonplanar massless two loop Feynman diagrams with four
  on-shell legs}},  \href{http://dx.doi.org/10.1016/S0370-2693(99)01277-0}{{\em
  Phys. Lett. B} {\bfseries 469} (1999) 225--234}
  [\href{http://arxiv.org/abs/hep-ph/9909506}{{\ttfamily hep-ph/9909506}}]
  [\href{http://inspirehep.net/search?p=find+Tausk:1999vh}{{\ttfamily
  InSPIRE}}].

\bibitem{Czakon:2005rk}
M.~Czakon, {\it {Automatized analytic continuation of Mellin-Barnes
  integrals}},  \href{http://dx.doi.org/10.1016/j.cpc.2006.07.002}{{\em Comput.
  Phys. Commun.} {\bfseries 175} (2006) 559--571}
  [\href{http://arxiv.org/abs/hep-ph/0511200}{{\ttfamily hep-ph/0511200}}]
  [\href{http://inspirehep.net/search?p=find+Czakon:2005rk}{{\ttfamily
  InSPIRE}}].

\bibitem{Smirnov:2009up}
A.~V. Smirnov and V.~A. Smirnov, {\it {On the Resolution of Singularities of
  Multiple Mellin-Barnes Integrals}},
  \href{http://dx.doi.org/10.1140/epjc/s10052-009-1039-6}{{\em Eur. Phys. J. C}
  {\bfseries 62} (2009) 445--449}
  [\href{http://arxiv.org/abs/0901.0386}{{\ttfamily arXiv:0901.0386}}]
  [\href{http://inspirehep.net/search?p=find+Smirnov:2009up}{{\ttfamily
  InSPIRE}}].

\bibitem{Gluza:2007rt}
J.~Gluza, K.~Kajda, and T.~Riemann, {\it {AMBRE: A Mathematica package for the
  construction of Mellin-Barnes representations for Feynman integrals}},
  \href{http://dx.doi.org/10.1016/j.cpc.2007.07.001}{{\em Comput. Phys.
  Commun.} {\bfseries 177} (2007) 879--893}
  [\href{http://arxiv.org/abs/0704.2423}{{\ttfamily arXiv:0704.2423}}]
  [\href{http://inspirehep.net/search?p=find+Gluza:2007rt}{{\ttfamily
  InSPIRE}}].

\bibitem{Liu:2017jxz}
X.~Liu, Y.-Q. Ma, and C.-Y. Wang, {\it {A Systematic and Efficient Method to
  Compute Multi-loop Master Integrals}},
\href{http://dx.doi.org/10.1016/j.physletb.2018.02.026}{{\em Phys. Lett.}
  {\bfseries B779} (2018) 353--357}
  [\href{http://arxiv.org/abs/1711.09572}{{\ttfamily arXiv:1711.09572}}]
  [\href{http://inspirehep.net/search?p=find+Liu:2017jxz}{{\ttfamily
  InSPIRE}}].

\bibitem{Chetyrkin:1981qh}
K.~G. Chetyrkin and F.~V. Tkachov, {\it {Integration by Parts: The Algorithm to
  Calculate beta Functions in 4 Loops}},
\href{http://dx.doi.org/10.1016/0550-3213(81)90199-1}{{\em Nucl. Phys.}
  {\bfseries B192} (1981) 159--204}
  [\href{http://inspirehep.net/search?p=find+Chetyrkin:1981qh}{{\ttfamily
  InSPIRE}}].

\bibitem{Gluza:2010ws}
J.~Gluza, K.~Kajda, and D.~A. Kosower, {\it {Towards a Basis for Planar
  Two-Loop Integrals}},
\href{http://dx.doi.org/10.1103/PhysRevD.83.045012}{{\em Phys. Rev.} {\bfseries
  D83} (2011) 045012} [\href{http://arxiv.org/abs/1009.0472}{{\ttfamily
  arXiv:1009.0472}}]
  [\href{http://inspirehep.net/search?p=find+Gluza:2010ws}{{\ttfamily
  InSPIRE}}].

\bibitem{Schabinger:2011dz}
R.~M. Schabinger, {\it {A New Algorithm For The Generation Of
  Unitarity-Compatible Integration By Parts Relations}},
\href{http://dx.doi.org/10.1007/JHEP01(2012)077}{{\em JHEP} {\bfseries 01}
  (2012) 077} [\href{http://arxiv.org/abs/1111.4220}{{\ttfamily
  arXiv:1111.4220}}]
  [\href{http://inspirehep.net/search?p=find+Schabinger:2011dz}{{\ttfamily
  InSPIRE}}].

\bibitem{vonManteuffel:2012np}
A.~von Manteuffel and C.~Studerus,
{\it {Reduze 2 - Distributed Feynman Integral Reduction}},
  [\href{http://arxiv.org/abs/1201.4330}{{\ttfamily arXiv:1201.4330}}]
  [\href{http://inspirehep.net/search?p=find+vonManteuffel:2012np}{{\ttfamily
  InSPIRE}}].

\bibitem{Lee:2013mka}
R.~N. Lee, {\it {LiteRed 1.4: a powerful tool for reduction of multiloop
  integrals}},
\href{http://dx.doi.org/10.1088/1742-6596/523/1/012059}{{\em J. Phys. Conf.
  Ser.} {\bfseries 523} (2014) 012059}
  [\href{http://arxiv.org/abs/1310.1145}{{\ttfamily arXiv:1310.1145}}]
  [\href{http://inspirehep.net/search?p=find+Lee:2013mka}{{\ttfamily
  InSPIRE}}].

\bibitem{vonManteuffel:2014ixa}
A.~von Manteuffel and R.~M. Schabinger, {\it {A novel approach to integration
  by parts reduction}},
\href{http://dx.doi.org/10.1016/j.physletb.2015.03.029}{{\em Phys. Lett.}
  {\bfseries B744} (2015) 101--104}
  [\href{http://arxiv.org/abs/1406.4513}{{\ttfamily arXiv:1406.4513}}]
  [\href{http://inspirehep.net/search?p=find+vonManteuffel:2014ixa}{{\ttfamily
  InSPIRE}}].

\bibitem{Larsen:2015ped}
K.~J. Larsen and Y.~Zhang, {\it {Integration-by-parts reductions from unitarity
  cuts and algebraic geometry}},
\href{http://dx.doi.org/10.1103/PhysRevD.93.041701}{{\em Phys. Rev.} {\bfseries
  D93} (2016) 041701} [\href{http://arxiv.org/abs/1511.01071}{{\ttfamily
  arXiv:1511.01071}}]
  [\href{http://inspirehep.net/search?p=find+Larsen:2015ped}{{\ttfamily
  InSPIRE}}].

\bibitem{Peraro:2016wsq}
T.~Peraro, {\it {Scattering amplitudes over finite fields and multivariate
  functional reconstruction}},
\href{http://dx.doi.org/10.1007/JHEP12(2016)030}{{\em JHEP} {\bfseries 12}
  (2016) 030} [\href{http://arxiv.org/abs/1608.01902}{{\ttfamily
  arXiv:1608.01902}}]
  [\href{http://inspirehep.net/search?p=find+Peraro:2016wsq}{{\ttfamily
  InSPIRE}}].

\bibitem{Mastrolia:2018uzb}
P.~Mastrolia and S.~Mizera, {\it {Feynman Integrals and Intersection Theory}},
\href{http://dx.doi.org/10.1007/JHEP02(2019)139}{{\em JHEP} {\bfseries 02}
  (2019) 139} [\href{http://arxiv.org/abs/1810.03818}{{\ttfamily
  arXiv:1810.03818}}]
  [\href{http://inspirehep.net/search?p=find+Mastrolia:2018uzb}{{\ttfamily
  InSPIRE}}].

\bibitem{Liu:2018dmc}
X.~Liu and Y.-Q. Ma, {\it {Determining arbitrary Feynman integrals by vacuum
  integrals}},  \href{http://dx.doi.org/10.1103/PhysRevD.99.071501}{{\em Phys.
  Rev. D} {\bfseries 99} (2019) 071501}
  [\href{http://arxiv.org/abs/1801.10523}{{\ttfamily arXiv:1801.10523}}]
  [\href{http://inspirehep.net/search?p=find+Liu:2018dmc}{{\ttfamily
  InSPIRE}}].

\bibitem{Guan:2019bcx}
X.~Guan, X.~Liu, and Y.-Q. Ma, {\it {Complete reduction of integrals in
  two-loop five-light-parton scattering amplitudes}},
  \href{http://dx.doi.org/10.1088/1674-1137/44/9/093106}{{\em Chin. Phys. C}
  {\bfseries 44} (2020) 093106}
  [\href{http://arxiv.org/abs/1912.09294}{{\ttfamily arXiv:1912.09294}}]
  [\href{http://inspirehep.net/search?p=find+Guan:2019bcx}{{\ttfamily
  InSPIRE}}].

\bibitem{Klappert:2019emp}
J.~Klappert and F.~Lange, {\it {Reconstructing rational functions with
  FireFly}},  \href{http://dx.doi.org/10.1016/j.cpc.2019.106951}{{\em Comput.
  Phys. Commun.} {\bfseries 247} (2020) 106951}
  [\href{http://arxiv.org/abs/1904.00009}{{\ttfamily arXiv:1904.00009}}]
  [\href{http://inspirehep.net/search?p=find+Klappert:2019emp}{{\ttfamily
  InSPIRE}}].

\bibitem{Peraro:2019svx}
T.~Peraro, {\it {FiniteFlow: multivariate functional reconstruction using
  finite fields and dataflow graphs}},
\href{http://dx.doi.org/10.1007/JHEP07(2019)031}{{\em JHEP} {\bfseries 07}
  (2019) 031} [\href{http://arxiv.org/abs/1905.08019}{{\ttfamily
  arXiv:1905.08019}}]
  [\href{http://inspirehep.net/search?p=find+Peraro:2019svx}{{\ttfamily
  InSPIRE}}].

\bibitem{Frellesvig:2019kgj}
H.~Frellesvig, F.~Gasparotto, S.~Laporta, M.~K. Mandal, P.~Mastrolia,
  L.~Mattiazzi, and S.~Mizera, {\it {Decomposition of Feynman Integrals on the
  Maximal Cut by Intersection Numbers}},
\href{http://dx.doi.org/10.1007/JHEP05(2019)153}{{\em JHEP} {\bfseries 05}
  (2019) 153} [\href{http://arxiv.org/abs/1901.11510}{{\ttfamily
  arXiv:1901.11510}}]
  [\href{http://inspirehep.net/search?p=find+Frellesvig:2019kgj}{{\ttfamily
  InSPIRE}}].

\bibitem{Wang:2019mnn}
Y.~Wang, Z.~Li, and N.~Ul~Basat, {\it {Direct reduction of multiloop multiscale
  scattering amplitudes}},
  \href{http://dx.doi.org/10.1103/PhysRevD.101.076023}{{\em Phys. Rev. D}
  {\bfseries 101} (2020) 076023}
  [\href{http://arxiv.org/abs/1901.09390}{{\ttfamily arXiv:1901.09390}}]
  [\href{http://inspirehep.net/search?p=find+Wang:2019mnn}{{\ttfamily
  InSPIRE}}].

\bibitem{Smirnov:2019qkx}
A.~V. Smirnov and F.~S. Chuharev, {\it {FIRE6: Feynman Integral REduction with
  Modular Arithmetic}},
  \href{http://dx.doi.org/10.1016/j.cpc.2019.106877}{{\em Comput. Phys.
  Commun.} {\bfseries 247} (2020) 106877}
  [\href{http://arxiv.org/abs/1901.07808}{{\ttfamily arXiv:1901.07808}}]
  [\href{http://inspirehep.net/search?p=find+Smirnov:2019qkx}{{\ttfamily
  InSPIRE}}].

\bibitem{Klappert:2020nbg}
J.~Klappert, F.~Lange, P.~Maierh\"ofer, and J.~Usovitsch, {\it {Integral
  reduction with Kira 2.0 and finite field methods}},
  \href{http://dx.doi.org/10.1016/j.cpc.2021.108024}{{\em Comput. Phys.
  Commun.} {\bfseries 266} (2021) 108024}
  [\href{http://arxiv.org/abs/2008.06494}{{\ttfamily arXiv:2008.06494}}]
  [\href{http://inspirehep.net/search?p=find+Klappert:2020nbg}{{\ttfamily
  InSPIRE}}].

\bibitem{Boehm:2020ijp}
J.~Boehm, M.~Wittmann, Z.~Wu, Y.~Xu, and Y.~Zhang, {\it {IBP reduction
  coefficients made simple}},
  \href{http://dx.doi.org/10.1007/JHEP12(2020)054}{{\em JHEP} {\bfseries 12}
  (2020) 054} [\href{http://arxiv.org/abs/2008.13194}{{\ttfamily
  arXiv:2008.13194}}]
  [\href{http://inspirehep.net/search?p=find+Boehm:2020ijp}{{\ttfamily
  InSPIRE}}].

\bibitem{Heller:2021qkz}
M.~Heller and A.~von Manteuffel, {\it {MultivariateApart: Generalized partial
  fractions}},  \href{http://dx.doi.org/10.1016/j.cpc.2021.108174}{{\em Comput.
  Phys. Commun.} {\bfseries 271} (2022) 108174}
  [\href{http://arxiv.org/abs/2101.08283}{{\ttfamily arXiv:2101.08283}}]
  [\href{http://inspirehep.net/search?p=find+Heller:2021qkz}{{\ttfamily
  InSPIRE}}].

\bibitem{Bendle:2021ueg}
D.~Bendle, J.~Boehm, M.~Heymann, R.~Ma, M.~Rahn, L.~Ristau, M.~Wittmann, Z.~Wu,
  and Y.~Zhang, {\it {Two-loop five-point integration-by-parts relations in a
  usable form}},  [\href{http://arxiv.org/abs/2104.06866}{{\ttfamily
  arXiv:2104.06866}}]
  [\href{http://inspirehep.net/search?p=find+Bendle:2021ueg}{{\ttfamily
  InSPIRE}}].

\bibitem{Song:2021vru}
Q.~Song and A.~Freitas, {\it {On the evaluation of two-loop electroweak box
  diagrams for $e^+e^- \to HZ$ production}},
  \href{http://dx.doi.org/10.1007/JHEP04(2021)179}{{\em JHEP} {\bfseries 04}
  (2021) 179} [\href{http://arxiv.org/abs/2101.00308}{{\ttfamily
  arXiv:2101.00308}}]
  [\href{http://inspirehep.net/search?p=find+Song:2021vru}{{\ttfamily
  InSPIRE}}].

\bibitem{Chicherin:2018old}
D.~Chicherin, T.~Gehrmann, J.~M. Henn, P.~Wasser, Y.~Zhang, and S.~Zoia, {\it
  {All Master Integrals for Three-Jet Production at Next-to-Next-to-Leading
  Order}},  \href{http://dx.doi.org/10.1103/PhysRevLett.123.041603}{{\em Phys.
  Rev. Lett.} {\bfseries 123} (2019) 041603}
  [\href{http://arxiv.org/abs/1812.11160}{{\ttfamily arXiv:1812.11160}}]
  [\href{http://inspirehep.net/search?p=find+Chicherin:2018old}{{\ttfamily
  InSPIRE}}].

\bibitem{Papadopoulos:2015jft}
C.~G. Papadopoulos, D.~Tommasini, and C.~Wever, {\it {The Pentabox Master
  Integrals with the Simplified Differential Equations approach}},
  \href{http://dx.doi.org/10.1007/JHEP04(2016)078}{{\em JHEP} {\bfseries 04}
  (2016) 078} [\href{http://arxiv.org/abs/1511.09404}{{\ttfamily
  arXiv:1511.09404}}]
  [\href{http://inspirehep.net/search?p=find+Papadopoulos:2015jft}{{\ttfamily
  InSPIRE}}].

\bibitem{Gehrmann:2018yef}
T.~Gehrmann, J.~M. Henn, and N.~A. Lo~Presti, {\it {Pentagon functions for
  massless planar scattering amplitudes}},
\href{http://dx.doi.org/10.1007/JHEP10(2018)103}{{\em JHEP} {\bfseries 10}
  (2018) 103} [\href{http://arxiv.org/abs/1807.09812}{{\ttfamily
  arXiv:1807.09812}}]
  [\href{http://inspirehep.net/search?p=find+Gehrmann:2018yef}{{\ttfamily
  InSPIRE}}].

\bibitem{Chicherin:2018mue}
D.~Chicherin, T.~Gehrmann, J.~M. Henn, N.~A. Lo~Presti, V.~Mitev, and
  P.~Wasser, {\it {Analytic result for the nonplanar hexa-box integrals}},
  \href{http://dx.doi.org/10.1007/JHEP03(2019)042}{{\em JHEP} {\bfseries 03}
  (2019) 042} [\href{http://arxiv.org/abs/1809.06240}{{\ttfamily
  arXiv:1809.06240}}]
  [\href{http://inspirehep.net/search?p=find+Chicherin:2018mue}{{\ttfamily
  InSPIRE}}].

\bibitem{Chicherin:2020oor}
D.~Chicherin and V.~Sotnikov, {\it {Pentagon Functions for Scattering of Five
  Massless Particles}},  \href{http://dx.doi.org/10.1007/JHEP12(2020)167}{{\em
  JHEP} {\bfseries 12} (2020) 167}
  [\href{http://arxiv.org/abs/2009.07803}{{\ttfamily arXiv:2009.07803}}]
  [\href{http://inspirehep.net/search?p=find+Chicherin:2020oor}{{\ttfamily
  InSPIRE}}].

\bibitem{Abreu:2020jxa}
S.~Abreu, H.~Ita, F.~Moriello, B.~Page, W.~Tschernow, and M.~Zeng, {\it
  {Two-Loop Integrals for Planar Five-Point One-Mass Processes}},
  \href{http://dx.doi.org/10.1007/JHEP11(2020)117}{{\em JHEP} {\bfseries 11}
  (2020) 117} [\href{http://arxiv.org/abs/2005.04195}{{\ttfamily
  arXiv:2005.04195}}]
  [\href{http://inspirehep.net/search?p=find+Abreu:2020jxa}{{\ttfamily
  InSPIRE}}].

\bibitem{Canko:2020ylt}
D.~D. Canko, C.~G. Papadopoulos, and N.~Syrrakos, {\it {Analytic representation
  of all planar two-loop five-point Master Integrals with one off-shell leg}},
  \href{http://dx.doi.org/10.1007/JHEP01(2021)199}{{\em JHEP} {\bfseries 01}
  (2021) 199} [\href{http://arxiv.org/abs/2009.13917}{{\ttfamily
  arXiv:2009.13917}}]
  [\href{http://inspirehep.net/search?p=find+Canko:2020ylt}{{\ttfamily
  InSPIRE}}].

\bibitem{Davydychev:1992mt}
A.~I. Davydychev and J.~B. Tausk, {\it {Two loop selfenergy diagrams with
  different masses and the momentum expansion}},
\href{http://dx.doi.org/10.1016/0550-3213(93)90338-P}{{\em Nucl. Phys.}
  {\bfseries B397} (1993) 123--142}
  [\href{http://inspirehep.net/search?p=find+Davydychev:1992mt}{{\ttfamily
  InSPIRE}}].

\bibitem{Broadhurst:1998rz}
D.~J. Broadhurst, {\it {Massive three - loop Feynman diagrams reducible to SC*
  primitives of algebras of the sixth root of unity}},
\href{http://dx.doi.org/10.1007/s100529900935}{{\em Eur. Phys. J.} {\bfseries
  C8} (1999) 311--333} [\href{http://arxiv.org/abs/hep-th/9803091}{{\ttfamily
  hep-th/9803091}}]
  [\href{http://inspirehep.net/search?p=find+Broadhurst:1998rz}{{\ttfamily
  InSPIRE}}].

\bibitem{Schroder:2005va}
Y.~Schr\"{o}der and A.~Vuorinen, {\it {High-precision epsilon expansions of
  single-mass-scale four-loop vacuum bubbles}},
\href{http://dx.doi.org/10.1088/1126-6708/2005/06/051}{{\em JHEP} {\bfseries
  06} (2005) 051} [\href{http://arxiv.org/abs/hep-ph/0503209}{{\ttfamily
  hep-ph/0503209}}]
  [\href{http://inspirehep.net/search?p=find+Schroder:2005va}{{\ttfamily
  InSPIRE}}].

\bibitem{Luthe:2015ngq}
T.~Luthe, {\em {Fully massive vacuum integrals at 5 loops}}.
\newblock PhD thesis, Bielefeld U.,
\newblock 2015
  [\href{http://inspirehep.net/search?p=find+Luthe:2015ngq}{{\ttfamily
  InSPIRE}}].
\newblock
\url{https://pub.uni-bielefeld.de/publication/2776013}.
\newblock

\bibitem{Kniehl:2017ikj}
B.~A. Kniehl, A.~F. Pikelner, and O.~L. Veretin, {\it {Three-loop massive
  tadpoles and polylogarithms through weight six}},
\href{http://dx.doi.org/10.1007/JHEP08(2017)024}{{\em JHEP} {\bfseries 08}
  (2017) 024} [\href{http://arxiv.org/abs/1705.05136}{{\ttfamily
  arXiv:1705.05136}}]
  [\href{http://inspirehep.net/search?p=find+Kniehl:2017ikj}{{\ttfamily
  InSPIRE}}].

\bibitem{Luthe:2017ttc}
T.~Luthe, A.~Maier, P.~Marquard, and Y.~Schroder, {\it {Complete
  renormalization of QCD at five loops}},
\href{http://dx.doi.org/10.1007/JHEP03(2017)020}{{\em JHEP} {\bfseries 03}
  (2017) 020} [\href{http://arxiv.org/abs/1701.07068}{{\ttfamily
  arXiv:1701.07068}}]
  [\href{http://inspirehep.net/search?p=find+Luthe:2017ttc}{{\ttfamily
  InSPIRE}}].

\bibitem{Bronnum-Hansen:2020mzk}
C.~Br\o{}nnum-Hansen and C.-Y. Wang, {\it {Contribution of third generation
  quarks to two-loop helicity amplitudes for W boson pair production in gluon
  fusion}},  \href{http://dx.doi.org/10.1007/JHEP01(2021)170}{{\em JHEP}
  {\bfseries 01} (2021) 170} [\href{http://arxiv.org/abs/2009.03742}{{\ttfamily
  arXiv:2009.03742}}]
  [\href{http://inspirehep.net/search?p=find+Bronnum-Hansen:2020mzk}{{\ttfamily
  InSPIRE}}].

\bibitem{Bronnum-Hansen:2021olh}
C.~Br\o{}nnum-Hansen and C.-Y. Wang, {\it {Top quark contribution to two-loop
  helicity amplitudes for $Z$ boson pair production in gluon fusion}},
  \href{http://dx.doi.org/10.1007/JHEP05(2021)244}{{\em JHEP} {\bfseries 05}
  (2021) 244} [\href{http://arxiv.org/abs/2101.12095}{{\ttfamily
  arXiv:2101.12095}}]
  [\href{http://inspirehep.net/search?p=find+Bronnum-Hansen:2021olh}{{\ttfamily
  InSPIRE}}].

\bibitem{Beneke:1997zp}
M.~Beneke and V.~A. Smirnov, {\it {Asymptotic expansion of Feynman integrals
  near threshold}},
\href{http://dx.doi.org/10.1016/S0550-3213(98)00138-2}{{\em Nucl.Phys.}
  {\bfseries B522} (1998) 321--344}
  [\href{http://arxiv.org/abs/hep-ph/9711391}{{\ttfamily hep-ph/9711391}}]
  [\href{http://inspirehep.net/search?p=find+Beneke:1997zp}{{\ttfamily
  InSPIRE}}].

\bibitem{Smirnov:1999bza}
V.~A. Smirnov, {\it {Problems of the strategy of regions}},
  \href{http://dx.doi.org/10.1016/S0370-2693(99)01061-8}{{\em Phys. Lett. B}
  {\bfseries 465} (1999) 226--234}
  [\href{http://arxiv.org/abs/hep-ph/9907471}{{\ttfamily hep-ph/9907471}}]
  [\href{http://inspirehep.net/search?p=find+Smirnov:1999bza}{{\ttfamily
  InSPIRE}}].

\bibitem{Georgoudis:2021onj}
A.~Georgoudis, V.~Gon\c{c}alves, E.~Panzer, R.~Pereira, A.~V. Smirnov, and
  V.~A. Smirnov, {\it {Glue-and-cut at five loops}},
  \href{http://dx.doi.org/10.1007/JHEP09(2021)098}{{\em JHEP} {\bfseries 09}
  (2021) 098} [\href{http://arxiv.org/abs/2104.08272}{{\ttfamily
  arXiv:2104.08272}}]
  [\href{http://inspirehep.net/search?p=find+Georgoudis:2021onj}{{\ttfamily
  InSPIRE}}].

\bibitem{Liu:2022chg}
X.~Liu and Y.-Q. Ma, {\it {AMFlow: a Mathematica Package for Feynman integrals
  computation via Auxiliary Mass Flow}},
  [\href{http://arxiv.org/abs/2201.11669}{{\ttfamily arXiv:2201.11669}}]
  [\href{http://inspirehep.net/search?p=find+Liu:2022chg}{{\ttfamily
  InSPIRE}}].

\bibitem{Bonciani:2019jyb}
R.~Bonciani, V.~Del~Duca, H.~Frellesvig, J.~M. Henn, M.~Hidding, L.~Maestri,
  F.~Moriello, G.~Salvatori, and V.~A. Smirnov, {\it {Evaluating a family of
  two-loop non-planar master integrals for Higgs + jet production with full
  heavy-quark mass dependence}},
  \href{http://dx.doi.org/10.1007/JHEP01(2020)132}{{\em JHEP} {\bfseries 01}
  (2020) 132} [\href{http://arxiv.org/abs/1907.13156}{{\ttfamily
  arXiv:1907.13156}}]
  [\href{http://inspirehep.net/search?p=find+Bonciani:2019jyb}{{\ttfamily
  InSPIRE}}].

\bibitem{Frellesvig:2019byn}
H.~Frellesvig, M.~Hidding, L.~Maestri, F.~Moriello, and G.~Salvatori, {\it {The
  complete set of two-loop master integrals for Higgs + jet production in
  QCD}},  \href{http://dx.doi.org/10.1007/JHEP06(2020)093}{{\em JHEP}
  {\bfseries 06} (2020) 093} [\href{http://arxiv.org/abs/1911.06308}{{\ttfamily
  arXiv:1911.06308}}]
  [\href{http://inspirehep.net/search?p=find+Frellesvig:2019byn}{{\ttfamily
  InSPIRE}}].

\bibitem{Bailey:1999nv}
D.~H. Bailey and D.~J. Broadhurst, {\it {Parallel integer relation detection:
  Techniques and applications}},
  \href{http://dx.doi.org/10.1090/S0025-5718-00-01278-3}{{\em Math. Comput.}
  {\bfseries 70} (2001) 1719--1736}
  [\href{http://arxiv.org/abs/math/9905048}{{\ttfamily math/9905048}}]
  [\href{http://inspirehep.net/search?p=find+Bailey:1999nv}{{\ttfamily
  InSPIRE}}].

\end{thebibliography}
\end{document}